\documentclass[%
 reprint,
superscriptaddress,
 amsmath,amssymb,
 aps,
floatfix,
]{revtex4-2}

\usepackage{color}
\usepackage{graphicx}% Include figure files
\usepackage{dcolumn}% Align table columns on decimal point
\usepackage{bm}% bold math
\usepackage[varg]{txfonts}
\usepackage{hyperref}
\hypersetup{
    colorlinks=true,
    citecolor=blue,
    }
    
\usepackage{aas_macros} 

\newcommand{\modelname}{{ADNF}}

\begin{document}

\title{Detecting dense-matter phase transition signatures in neutron star mass-radius measurements as data anomalies using normalising flows}% Force line breaks with \\
%\thanks{}%

\author{Filip Morawski}
    \email{fmorawski@camk.edu.pl}
    \affiliation{Nicolaus Copernicus Astronomical Center, Polish Academy of Sciences, Bartycka 18, 00-716, Warsaw, Poland}
\author{Micha{\l} Bejger}
    \email{bejger@fe.infn.it}
    \affiliation{INFN Sezione di Ferrara, Via Saragat 1, 44122 Ferrara, Italy}
    \affiliation{Nicolaus Copernicus Astronomical Center, Polish Academy of Sciences, Bartycka 18, 00-716, Warsaw, Poland}

\date{\today}% It is always \today, today,

\begin{abstract}
Observations of neutron stars may be used to study aspects of extremely dense matter, specifically a possibility of phase transitions to exotic states, such as de-confined quarks. 

We present a novel data analysis method for detecting signatures of dense-matter phase transitions in sets of mass-radius measurements, and study its sensitivity with respect to the size of observational errors and the number of observations. The method is based on machine learning anomaly detection coupled with normalizing flows technique: the algorithm trained on samples of astrophysical observations featuring no phase transition signatures interprets a phase transition sample as an ``anomaly''. For the sake of this study, we focus on dense-matter equations of state leading to detached branches of mass-radius sequences (strong phase transitions), use an astrophysically-informed neutron-star mass function, and various magnitudes of observational errors and sample sizes. 

The method is shown to reliably detect cases of mass-radius relations with phase transition signatures, while increasing its sensitivity with decreasing measurement errors and increasing number of observations. We discuss marginal cases, when the phase transition mass is located near the edges of the mass function range. Evaluated on the current state-of-art selection of real measurements of electromagnetic and gravitational-wave observations, the method gives inconclusive results, which we interpret as due to small available sample size, large observational errors and complex systematics.
%\begin{description}
%\end{description}
\end{abstract}

 %Use showkeys class option if keyword
\maketitle

%________________________________________________________________
\section{Introduction}
\label{sect:intro} 
%________________________________________________________________

Neutron stars (NSs) are extremely dense and compact astrophysical objects, ideal for studying matter in conditions which  terrestrial experiments simply cannot replicate (see \cite{HaenselPY2007} for textbook introduction). In particular, NSs are used to uncover the details of the dense-matter equation of state (EOS) at densities several times higher than the nuclear saturation density $\rho_s\simeq 2.7\times 10^{14}\ \mathrm{g/cm^3}$ (baryon density $n_s{\simeq}\,0.16\,\mathrm{fm^{-3}}$). 

NSs are studied indirectly by means of astrophysical observations, either electromagnetic (EM) or using gravitational waves (GWs), therefore the EOS is recovered by measuring global NS features, i.e. global quantities such as the mass $M$, radius $R$ and tidal deformability $\Lambda$, related to the EOS by stellar structure equations, in the simplest case by the Tolman–Oppenheimer–Volkoff (TOV) equations \cite{Tolman1939,OppenheimerV1939} to which the EOS is an input. Recovering the EOS from the NS observations amounts to ``inverting'' the TOV equations. Because measurement uncertainties are an inseparable part of $M$, $R$ observations, recovery of the actual EOS values is a difficult task. Various EOSs may agree with the NS observations within the measurement error ranges. This ambiguity complicates studies of the interior of NSs, specifically in the assessment whether a dense matter phase transition to exotic matter (like the de-confinement to quark matter) indeed occurs; see \cite{PhysRevD.88.083013} for a generic phase diagram of possible forms of $M(R)$ relations for phase-transition EOS NSs, and \cite{2019JPhG...46k4001A} for a recent review of the subject.

These challenges motivate a novel method to search for the evidence for dense-matter phase transitions using EM observations of NSs, namely masses $M$ and radii $R$. The method is based on machine learning (ML) methodology, known as the anomaly detection (AD), i.e. searching for rare ``abnormal'' (``exotic'') events occurring sometimes in the otherwise ``standard'' data under study. Here, the abnormal (exotic) events will denote the collections of NS observations, associated with EOSs exhibiting phase transitions strong enough to be visible by means of these observations. The remaining data is associated with simpler EOSs that do not exhibit strong phase transitions (i.e. purely hadronic EOSs). The deviation from the expected ``standard'' behavior is quantified using the normalizing flow (NF) technique \cite{2019arXiv190809257K,papamakarios2021normalizing}. Henceforth, we will denote the anomaly detection normalizing flows model using the {\modelname} acronym.

{\parskip 1pt 
The project focuses on simulated observational data. We adopt parametric EOSs, as well as microphysically-motivated EOSs to study critical number of observations and the size of observational errors for which the {\modelname} model trained on NS datasets without phase transitions is able to recognise a collection of NS $M(R)$ observations obtained using a (sufficiently pronounced) phase transition EOS. In Sect. \ref{sect:methods}, we briefly outline the data generation procedure (Sect.~\ref{sect:meth_data}) and the ML algorithms used (Sect.~\ref{sect:meth_ml}). Section \ref{sect:results} contains results of the {\modelname} on simulated NS observation datasets, where we discuss the figures of merit and metrics used for the decision making. We conclude in Sect. \ref{sect:conclusions} with a summary and outlook. Additionally, Appendix~\ref{sect:real_data} contains a discussion of the evaluation on currently available real $M(R)$ measurements. 
}
%________________________________________________________________
\section{Methodology} 
\label{sect:methods} 
%________________________________________________________________

\subsection{Astrophysical data preparation} 
\label{sect:meth_data} 

The input data consists of two populations of EOSs: a dataset based on EOSs without strong phase transitions (as in \cite{morawski2020b}) used for both training and testing of the ADNF model as well as additional testing set with strong first-order phase transitions to quark phase (approximated by the Maxwell construction, as in \cite{2019A&A...622A.174S}), additionally supplied with an example of two families EOS scenario of \citeauthor{2014PhRvD..89d3014D} \cite{2014PhRvD..89d3014D}.
Masses and radii are obtained by solving the TOV equations, and on the basis of this sequences, training and testing data with simulated observational errors are obtained. Below we describe the procedure in detail. 

\begin{figure*}[ht]
    \includegraphics[height=0.7\columnwidth]{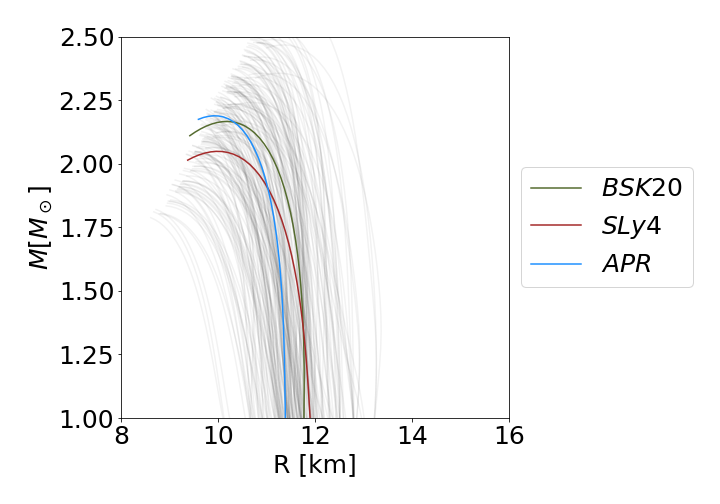}
    \includegraphics[height=0.7\columnwidth]{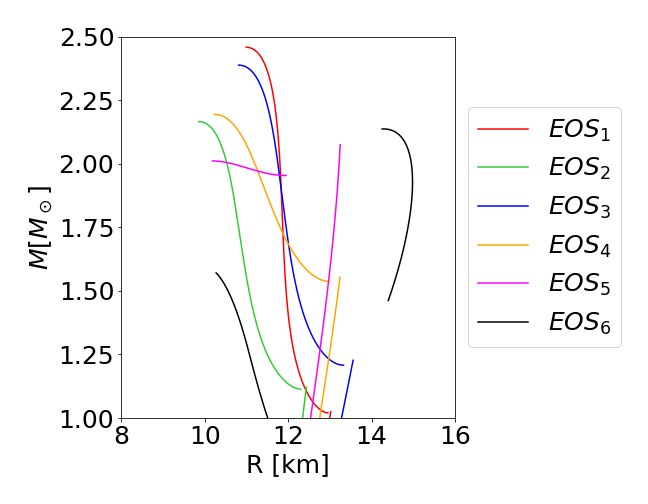}
    \caption{The mass-radius $M(R)$ relations generated using the piecewise relativistic polytrope models of EOSs, both without (\textit{left plot}, following \cite{morawski2020b}) and with phase transitions (\textit{right plot}, \cite{2019A&A...622A.174S} and \cite{2014PhRvD..89d3014D}). For comparison, on the left plot the reference SLy4 EOS \cite{HaenselP1994,DouchinH2001}, the APR EOS \cite{APR} and the BSK20 EOS \cite{BSK20} are included. On the right plot, the $EOS_{1-5}$ refer to relativistic polytrope connected to MIT-like bag quark EOS models with increasing average mass of the phase transition occurrence (see Table~\ref{tab:eos_pt} for details), whereas the $EOS_6$ denotes the $M(R)$ relation generated from the \cite{2014PhRvD..89d3014D} EOS.}%
    \label{fig:eos}
\end{figure*}

The first EOS training data set, denoted ``standard'' behavior is exhibiting no strong phase transitions, and is used to train the ML model. It is a parametric set of EOSs based on relativistic piecewise polytrope prescription for densities above the nuclear saturation density $\rho_s$, and a realistic crust EOS based on the SLy4 EOS \cite{HaenselP1994,DouchinH2001} for densities smaller than $\rho_s$. We use the set of EOSs of \cite{morawski2020b}; see their Table 1 for details. Left panel in Fig.~\ref{fig:eos} presents selected $M(R)$ relations computed for these EOSs, as well as three microphysically-motivated tabulated EOSs for comparison: the SLy4 EOS  \cite{HaenselP1994,DouchinH2001}, the APR EOS \cite{APR} and the BSK20 EOS \cite{BSK20}. The dataset results in $M(R)$ relations in widely accepted ranges consistent with current astrophysical observations (see Fig.~\ref{fig:mr-realdata}).

Second, the``abnormal'' (``exotic'') population used for testing purposes consists of six EOSs exhibiting strong phase transitions at different densities, and hence in different positions on the $M(R)$ plane. We aim here to demonstrate a ML method able to correctly detect the exotic $M(R)$ relations based on these EOS for phase transition onsets located at various $M(R)$ measurements, and yield a negative result if there is no chance of detecting the phase transition signature. 

Right panel in Fig.~\ref{fig:eos} presents $M(R)$ relations based on the EOSs, which construction is based on \cite{2019A&A...622A.174S}, employing the Maxwell construction between the ``normal'' phase approximated by a relativistic polytrope, and an ``exotic'' (quark) phase, approximated by MIT-bag like linear pressure-density EOS relation of \cite{2000A&A...359..311Z}. Specifically, $EOS_1-EOS_4$ feature pronounced softening leading to detached $M(R)$ branches at different values of $M$ and $R$, whereas $EOS_5$ mimics a $M(R)$ behavior of the so-called high-mass twin stars scenario \cite{2013arXiv1310.3803B,2017A&A...600A..39B}. The details of the $EOS_{1-5}$ are gathered in Table~\ref{tab:eos_pt}. 

Last but not least, $EOS_6$ belongs to a distinct class of two families scenario \cite{2014PhRvD..89d3014D}, which motivated by both the existence of massive NSs \cite{2010Natur.467.1081D,2013Sci...340..448A,2021ApJ...915L..12F} with large radii \cite{2021ApJ...918L..28M} and the necessity of the EOS softening at lower NS masses, resulting from postulated appearance of hyperons and $\Delta$ resonances \cite{2014PhRvD..89d3014D}; in this scenario, after crossing a critical density threshold, a nucleonic NS converts into quark star with a substantially larger radius, which allows reaching $2\,M_\odot$ and featuring a distinct separation of $M(R)$ branches. Using the taxonomy of \cite{PhysRevD.88.083013}, ``standard'' cases consist of EOSs without phase transitions and of the type C (connected), while the ``exotic'' cases are types B (both) and D (detached), featuring two distinct branches in the $M(R)$ plane.

\begin{table}
\caption{Details of EOSs approximating a strong first-order phase transition between a ``normal'' phase (relativistic polytrope with adiabatic index $\gamma_1$) and an ``exotic'' (quark) MIT-bag like model (linear pressure-density approximation of \cite{2000A&A...359..311Z}). The transition occurs at $n_1$ with a density jump of $\lambda=n_2/n_1$. The square of the speed of sound in a quark matter $\alpha=1$ for all the EOSs. The value of $n_0$ denotes a point at which the SLy4 crust \cite{HaenselP1994,DouchinH2001} is connected with the polytrope. For more details, see \cite{2019A&A...622A.174S}.}
\label{tab:eos_pt}
  %\centering
  \begin{ruledtabular} 
  \begin{tabular}{ccccc} 
    $EOS$ & $n_0$ [fm$^{-3}$] & $\gamma_1$ & $n_1$ [fm$^{-3}$] & $\lambda=n_2/n_1$ \\ \hline
    $EOS_1$ & 0.185 & 5.5 & 0.26  & 1.7  \\ 
    $EOS_2$ & 0.235 & 4.5 & 0.335 & 1.7  \\  
    $EOS_3$ & 0.16  & 5.0 & 0.26  & 1.8  \\
    $EOS_4$ & 0.185 & 4.5 & 0.335 & 1.7  \\  
    $EOS_5$ & 0.21  & 4.5 & 0.41  & 1.8  \\ 
  \end{tabular}
  \end{ruledtabular} 

\end{table}

For both sets, the procedure of obtaining the NS observations for training and testing is the same as in \cite{morawski2020b}. For each EOS the TOV equations are solved, resulting in a $M(R)$ sequence. To simulate a set of astrophysical observations, we randomly select $N$ measurements ($N$ equal to 10, 30 or 50 observations) using an observationally-informed probability distribution for known NS masses \cite{ns_masses}, namely a two component Gaussian mixture model, with mean values at $1.34\,M_{\odot}$ and $1.8\,M_{\odot}$, and standard deviations of ${\sim}0.07\,M_{\odot}$ and ${\sim}0.21\,M_{\odot}$ respectively. We set a lower edge of the mass distribution, $M_{min}=1\,M_\odot$. Additionally, we take into account the existence of observational errors. For each initial set of $N$ $M(R)$ points we select final $M$ and $R$ values from 3 normal probability distributions defined as follows: $\mathcal{N}(M, \sigma_{M_i})$, with $\sigma_{M_i}\in\{0.01\,M_\odot,\,0.05\,M_\odot,\,0.1\,M_\odot\}$, denoted as $\mathcal{N}_{M1}$, $\mathcal{N}_{M2}$ and $\mathcal{N}_{M3}$, respectively. For radii we have considered $\mathcal{N}(R, \sigma_{R_i})$, with $\sigma_{R_i}\in\{0.1\,\mathrm{km},\,0.5\,\mathrm{km},\,1\,\mathrm{km}\}$, denoted as $\mathcal{N}_{R1}$, $\mathcal{N}_{R2}$ and $\mathcal{N}_{R3}$, respectively. 

In total, the training dataset contains 15000 $M(R)$ sequences produced by solving TOV equations using parameterized EOSs. For each of these sequences we randomly selected $N_s$ $M(R)$ pairs, and thus the procedure is repeated $N_s=100$ times for each $M(R)$ sequence. As a result, each input EOS is represented at the training stage by $N_s$ different realisations of $N$ ``observations'' of $M(R)$, additionally subjected  to observational errors by drawing the error values from the probability distributions described above. This step is used in order to effectively increase the dataset size as the method described in Sect.~\ref{sect:meth_ml} requires sufficiently-large amount of data in order to properly learn. The datasets are split into 90\% and 10\% subsets for the training and model validation, respectively. 

To test the ADNF, 24000 $M(R)$ simulated observations were generated for every combination of $N$ and magnitude of observational errors, ($\mathcal{N}_M$, $\mathcal{N}_R$). The test dataset was divided evenly into ``standard'' and ``exotic'' instances, with the same data generation procedure used as with the training data. In the case of the ``exotic'' data, a dataset of 12000 samples consisted of 2000 $M(R)$ realizations for every of the studied EOSs with the phase transition. The other 12000 samples, on the other hand, contained realizations of $M(R)$ sequences with features similar to training data. Their appropriate EOSs were generated by choosing random parameters from Table 1 of \cite{morawski2020b}, and then the NS mass function and observational errors were applied. As a result, the final dataset, while similar to the training, contained different $M(R)$ observations. Many realization for different EOSs allowed to compute detection efficiencies of detecting anomalies associated with particular EOS, as described in Sect.~\ref{sect:results} and Tab. \ref{tab:detection_efficiency}.

%The same procedure is applied to the test data to obtain 12000 EOSs for both sets, i.e., the $M(R)$ sequences for EOSs with and without phase transitions. Generating $N_s$ realizations for every EOS allows to compute detection efficiencies of detecting anomalies associated with particular EOS, as described in Sect.~\ref{sect:results} and Tab. \ref{tab:detection_efficiency}.

\subsection{Machine learning methods} 
\label{sect:meth_ml} 

ML is a field of artificial intelligence (AI) concerning algorithms that are able to learn from the data without the need of being explicitly pre-programmed \cite{samuel_ml}. ML algorithms are capable of solving variety problems such as regression, classification or clustering. 

Here we employ a combination of two ML techniques, artificial neural networks (ANN) and normalizing flows (NF), for detecting signatures of dense-matter phase transition imprinted on the collections of $M(R)$ observables. We first briefly describe these learning methods and then we explain how they are applied to our problem.

ANNs \cite{rosenblatt1958perceptron} (for textbook review, see \cite{Goodfellow2016}) are mathematical models that are loosely based on neural networks found in brains. These algorithms use a network of connected nodes known as artificial neurons that can communicate with one another. Weights, which are parameters changed during the learning process, can be used to adjust the strength of connections. Before transmitting the signal further, each neuron may apply certain non-linear functions (called activation functions) to the sum of its inputs. Artificial neurons are aggregated into layers that, depending on the activation function used, may perform different transformations on their inputs. Complex ANNs composed of many neurons and various training algorithms (such as the backpropagation and stochastic gradient descent, see e.g. \cite{Goodfellow2016} and references therein) can capture complex non-linear relationships in data by composing hierarchical internal representations. The deeper (i.e., more complex) the algorithm, the more abstract features it can learn from the data.

\begin{figure*}
    \includegraphics[height=0.7\columnwidth]{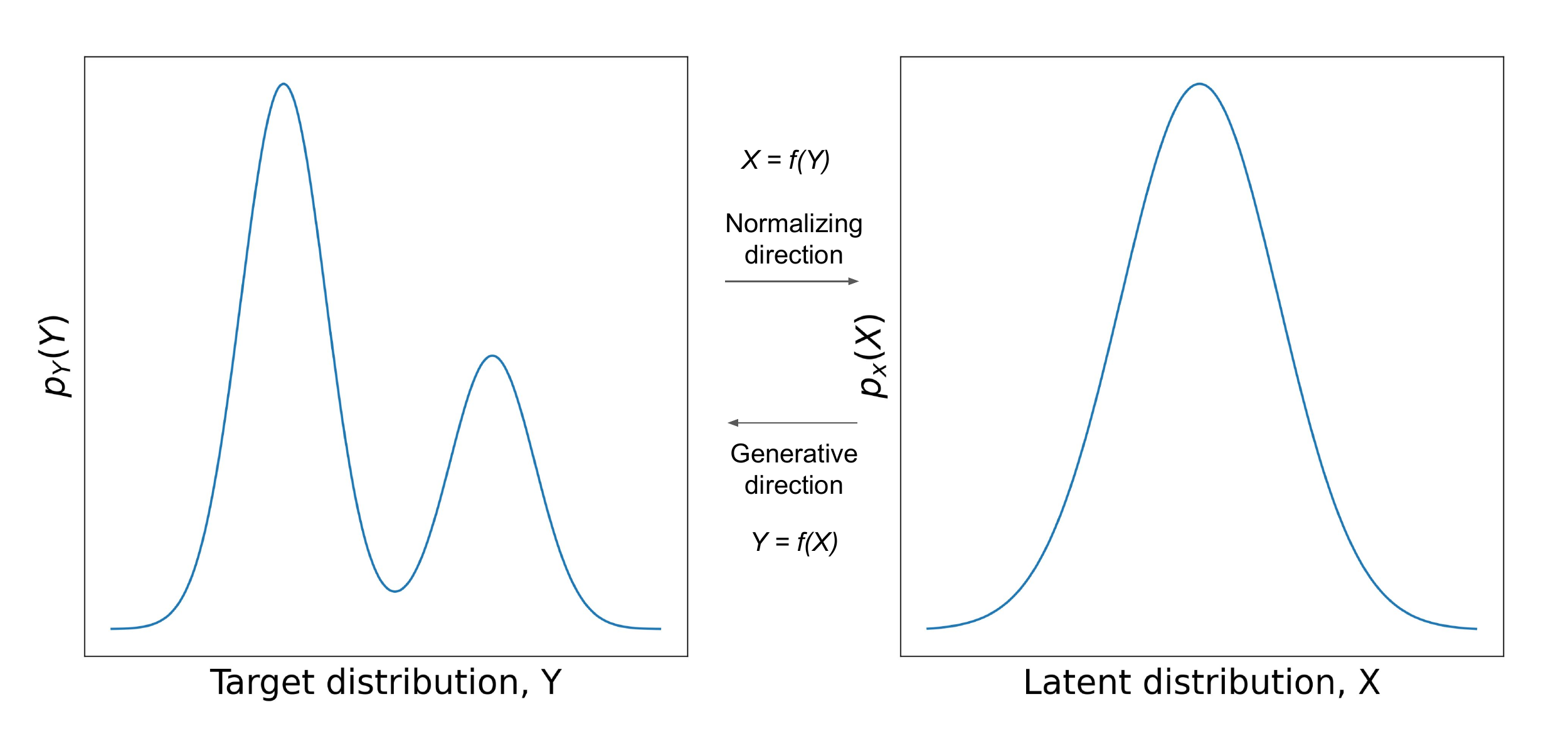}
    \caption{An illustration of invertible character of NF allowing change of variables and associated probability density functions via generation and normalization; see text for more details.}%
    \label{fig:nf_idea}
\end{figure*}

NFs belong to a family of generative models, which means they can learn the probability distribution from samples drawn from it. In the present context, the NF learns the underlying distributions of collections of $M(R)$ observables associated with different dense-matter EOSs. The NF is able not only to learn distribution parameters, but also generate new samples. Many generative models exist (see e.g. \cite{Kobyzev_2021} and references therein), but the majority of them does not provide a method for calculating the exact probability density for new samples; examples include the Variational Autoencoders (VAEs, see \cite{vae}) and Generative Adversarial Networks (GANs, see \cite{gan}).

NFs are reversible transformations of a simple distribution (e.g. standard normal) into a more complex distribution by a sequence of invertible and differentiable mappings. The simple distribution is referred to as a \textit{latent distribution} further in the text. The core principle allowing the transformation of variables following a certain distribution is the change of variables theorem \cite{changeOfVariables}, as shown in Fig.~\ref{fig:nf_idea}. According to it, the transformation for continuous, random variables, $X$ and $Y$, in $n$-dimensional space, related by mapping $f:\mathbb{R}^{n} \rightarrow \mathbb{R}^{n}$ such that $Y=f(X)$ and $X=f^{-1}(Y)$, is defined as:

\begin{equation}
    \begin{split}
    p_Y(Y) = p_X(f^{-1}(Y))\ \Big\lvert \det\left(\frac{\partial f^{-1}(Y)}{\partial Y}\right)\Big\rvert  = \\
    p_X(X)\ \Big\lvert\det\left(\frac{\partial f(X)}{\partial X}\right) \Big\rvert^{-1}.
    \end{split}
\end{equation}

The ${\partial f^{-1}(Y)}/{\partial Y}$ term is a square matrix of $n \times n$ dimensions known as the Jacobian matrix, which defines whether a transformation is invertible: only $\det({\partial f^{-1}(Y)}/{\partial Y}) \neq 0$ allows the inversion. Furthermore, if the determinant of the Jacobian is equal to unity, the mapping preserves its volume allowing $X, Y$ variables to have the same dimension.

In practical terms, input variables go through a chain of invertible transformations, where they are repeatedly substituted for new ones, eventually leading to a probability distribution of the final target variable. The density $p_L(Y)$ obtained by transforming a random variable $Y_0$ with distribution $p_0(Y)$ through a chain of L transformations $f_l$ is:

\begin{equation}
    \begin{split}
    \ln p_L(Y_L) = \ln p_0(Y_0) - \sum_{l=1}^{L} \ln \Big\lvert \det\left( \frac{\partial f_l (Y_{l-1})}{\partial Y_{l-1}}\right) \Big\rvert.
    \end{split}
\end{equation}

The path that the initial variable $Y_0$ travels throughout the chain of transformation is called the \textit{flow}, while the path formed by distributions $p_l$ is called the \textit{normalizing flow}, hence the name of the method. 

%Different implementations of flows allow for the transformation from a complex distribution into a simple, latent one. However, the one used in the presented work was based on the \textit{affine coupling} \cite{nice}. According to this method, the input data is split into two parts, where the first $n$ dimensions stay the same, whereas the second $n + 1$ to $N$ dimensions are updated conditioned on values of the first unchanged part of the vector which are being put through the ANN to learn the parameters of the transformation.
Various flow implementations enable the transformation of a complex distribution into a simple, latent one. The one used in this work is based on the \textit{affine coupling} \cite{nice}. According to that method, the input data is divided into two parts. The first $n$ dimensions remain the same. On the other hand, the second $n + 1$ to $N$ dimensions go through an affine transformation (defined in terms of a scale and translation operations), with the parameters for this transformation learned using the first part of the data that goes through the ANN.

More expressive mapping can be achieved by stacking multiple coupling transforms and alternating which part of the input vector is updated. Here, the generative aspect of NFs was used in a limited capacity to convert the simulated NS observations to latent representation distributions, in order to study the differences in the latent distributions produced by both ``standard'' and ``exotic'' datasets, on the premise that abnormal (exotic) events produce significantly different latent distributions than the standard events.

\begin{figure}[ht]
    %\centering
    \includegraphics[width=0.5\textwidth]{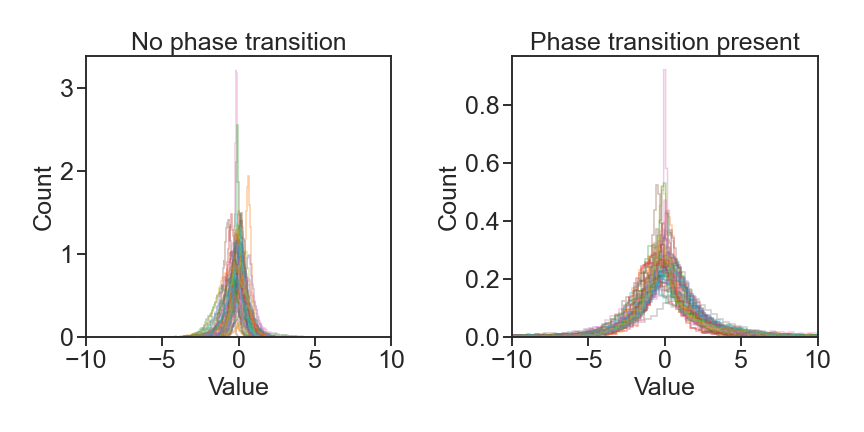}
    \vspace{0.1ex}  
    \includegraphics[width=0.5\textwidth]{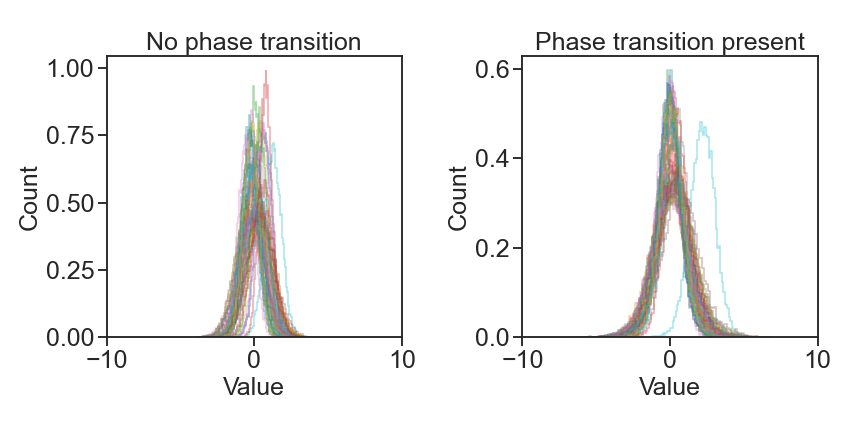}
    \caption{The {\modelname} latent representation for the case of training on $N=30$ $M(R)$ data points without measurement uncertainties (\textit{top row}) and with uncertainties drawn from $\mathcal{N}_{M3}$ and $\mathcal{N}_{R3}$ (\textit{bottom row}), for latent space size $K\equiv 2N=60$. For the ``exotic'' data (\textit{right panels}), certain deviations in the latent representation appear with respect to the ``standard'' case (\textit{left panels}). In the \textit{bottom plots} the deviations are smaller because observational errors were applied.}
    \label{fig:latent_representation}
\end{figure}

The final architecture of the NF was chosen after a set of empirical tests for all combinations of number of observations and measurement uncertainties. The final {\modelname} model consisted of 4 transformations, 4 layers per transformations and 4 neurons per layers. We chose the ADAM optimizer to train the NF with learning rate equal to $10^{-4}$ and the weight decay equal to $10^{-6}$. The model was trained for 300 epochs with a batch size of 1024 (see the definition of parameters in e.g. \cite{Goodfellow2016}). The implementation was carried out using the {\tt Python} \cite{10.5555/1593511} {\tt NFlows} library \cite{nflows} on top of the {\tt PyTorch} library \cite{pytorch} with support for the GPU.

%________________________________________________________________
\section{Results} 
\label{sect:results} 
%________________________________________________________________

The {\modelname} model was trained on datasets with varying numbers of observations and measurement uncertainties, described in Sect.~\ref{sect:meth_data}. We will start with describing the latent representation resulting from the analysis. 

As described in Sect.~\ref{sect:meth_ml}, NF transforms the input data into a simple distribution (e.g. standard normal) according to the transformation of variables theorem. Since the ``exotic'' dataset contains $M(R)$ observations associated with EOS distinct from the ``standard'' data used for the training of our model, we expected a different outcomes of the mentioned transformation. In the first stage of our analysis, we studied the deviations of the latent representation between the two datasets.

First, we studied deviations in the latent representation of the {\modelname} for different configurations of data in terms of number of observations, measurement uncertainties and the presence of the phase transition in the collections of $M(R)$ observations. An example is shown in Fig.~\ref{fig:latent_representation} for the case of the number of observations $N=30$, for the NS observables generated for EOSs without a phase transition (left column) and with a phase transition (right column). Top two plots present the results of training on dataset without measurement uncertainties whereas the bottom panels with errors from $\mathcal{N}_{M3}$ and $\mathcal{N}_{R3}$. On all plots the number of distributions correspond to the dimensionality of the input data $K\equiv 2N$, i.e. $N$ concatenated pairs of NS masses and radii values; in the presented cases $N=30$, so the number of distributions is $K=60$. The differences between the ``standard'' and ``exotic'' cases are most pronounced in the top row, where measurement uncertainties were not taken into account. Even for the largest of the considered observational errors, however, some differences in the latent representation are visible, indicating that the {\modelname} could indeed transform the initial data to a distribution defined by different features.

\begin{figure*}
    \includegraphics[width=\textwidth]{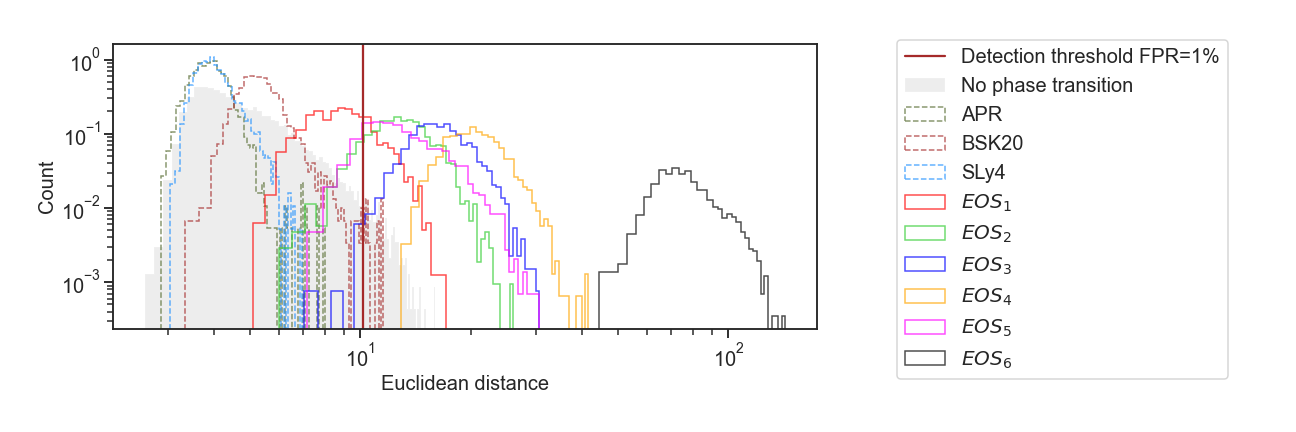}
    \vspace{-2ex}  
    \includegraphics[width=\textwidth]{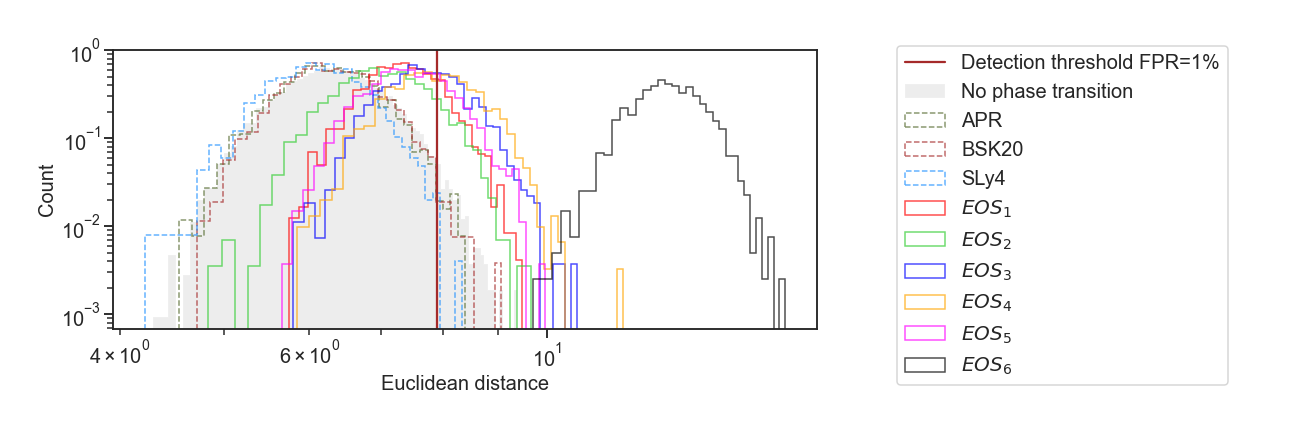}
    \caption{Distribution of Euclidean distances D (Eq.~\ref{eq:eucl_distance}) of NF latent representation for EOSs without phase transition (filled gray histogram denoting the dataset with no phase transition with features known from the training, dashed line histograms for the APR, SLy4 and BSK20 EOSs added for comparison), and with phase transition (colored, solid line histograms). \textit{Top plot}: results for the {\modelname} trained on data with no measurement uncertainties. \textit{Bottom plot}: results for model trained on data with $\mathcal{N}_{M3}$ and $\mathcal{N}_{R3}$ uncertainties. In both cases number of observations $N=30$ (dimensionality of latent representation $K=60$). Additionally, a brown vertical line represents a pre-selected anomaly detection threshold of False Positive Rate FPR=$1\%$. The microphysical models (APR, SLy4 and BSK20 EOSs) overlap with training samples to a very high extent, confirming that {\modelname} learned during training features of observables related to the lack of phase transition in the EOS.}
    %proper training (in terms of features associated with lack of the phase transition in the EOS).}
    \label{fig:euclidean_distance}
\end{figure*}

Since manual comparisons of high-dimensional space of latent representations are impracticable, we computed the Euclidean distances $D$ from the center of $K$-dimensional space for each realization of NS observables, by introducing 

\begin{equation}
    D = \sqrt{\sum_{i=1}^{K}\,x_{i}^2},
    \label{eq:eucl_distance} 
\end{equation} 

where $x_i$ refers to the $i$-th observation from the collection of $K=2N$ points, with $N$ being the number of $M(R)$ observations. As a result, a set of $N$ $M(R)$ observables were defined by a single value simplifying the further analysis. Computed Euclidean distances were then used to compare different $M(R)$ relations stemming from EOS with phase transition, as well as samples without phase transitions. The results are shown in Fig. \ref{fig:euclidean_distance}, where for $N=30$ ($K=60$), the top plot represents results for the {\modelname} trained on a dataset without measurement uncertainties, and the bottom plot results for the training on data with $\mathcal{N}_{M3}$ and $\mathcal{N}_{R3}$ uncertainties.

\begin{figure}
    \includegraphics[width=\columnwidth]{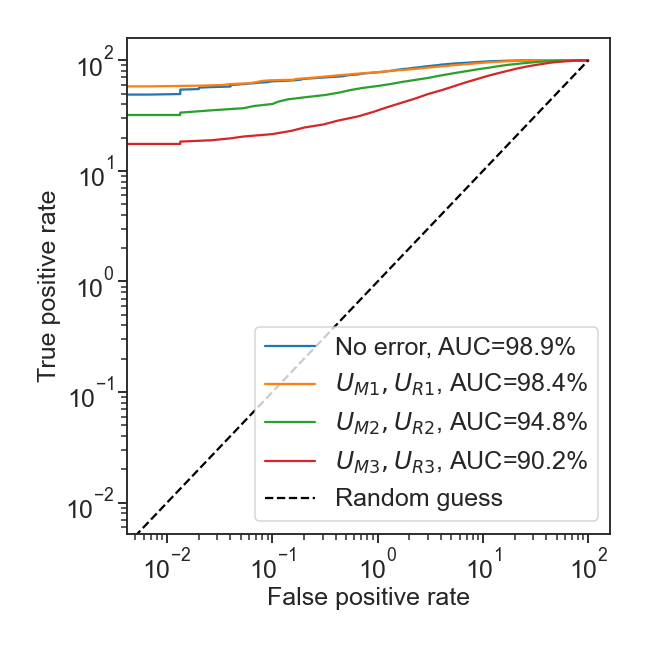}
    \caption{The ROC curves of the {\modelname} for $N=30$ ($K=60$) trained on all four cases of measurement uncertainty (no errors, as well ass $\mathcal{N}_{M1-3}$ and $\mathcal{N}_{R1-3}$ uncertainties). Value of the AUC is added to additionally present the effect of increasing measurement uncertainties on the performance of the {\modelname}.}
    \label{fig:roc}
\end{figure}

The histogram corresponding to the dataset without phase transition is on the left-most side of both plots, around the smallest values of Euclidean distance. This result is to be expected because this type of data is familiar from the training. In other cases, the exotic $M(R)$ measurements have higher Euclidean distance values. 
%This pattern confirms our intend, as the anomalous events in the latent representation should be placed ``further away'' from $\mathcal{N}(0,1)$, i.e. their Euclidean distance should be larger. 
This pattern agreed with our expectation, as the ``exotic'' instances in the latent representation should deviatee from the ``standard'' case.  
When we include measurement uncertainties in the data, the pattern changes in a predictable way. With increasing error size, for the majority of test EOSs, histograms begin to resemble more and more the case without phase transitions, which is to be expected as the data becomes more noisy, or the deviations in the latent representation decrease.
The vertical line on the histograms represents a detection threshold for anomaly detection, determined using the Receiver-Operating-Characteristic (ROC) curves shown in Fig.~\ref{fig:roc}. 
To compute the ROC we considered all of the ``exotic'' EOS $M(R)$ instances as one class - \textit{positive}, while the ``standard'' fell into the \textit{negative} class, which was necessary to define true positive rates (TPR) and false positive rates (FPR). Then, TPR and FPR were calculated by varying the Euclidean distance and counting the number of correctly detected events above the threshold. 
%(positive refer to the data with phase transition whereas negative to cases without phase transition).
Larger measurement uncertainties, as shown in Fig.~\ref{fig:roc}, lead to the worse detection capabilities of the {\modelname} defined in terms of the Area-Under-Curve (AUC). In this study, we set the anomaly detection threshold to FPR=$1\%$. 

Table~\ref{tab:auc} contains a comparison of AUC for all studied cases of numbers of observations and measurement uncertainties. Measurement uncertainties had the greatest impact on the {\modelname}'s performance. The greater the error in NS observables, the poorer the {\modelname} detection capabilities. Increasing the number of observations $N$ compensates partially the increased error size.

\begin{table}
\caption{The AUC of ROC curves for all cases of measurement uncertainty and number of observations included in this study. As one can see, the AUC decreases as uncertainties increase, but increases with increasing $N$. Increasing the number of observations $N$ compensates partially the increased error size.}
\label{tab:auc}
  %\centering
  %\resizebox{\columnwidth}{!}{
  \begin{ruledtabular}
  \begin{tabular}{cccc} 
    $M(R)$ & $N = 10$ & $N  = 30$ & $N  = 50$
    \\ \hline
    No error & 97.3\% & 98.9\% & 99.2\% \\ 
    $\mathcal{N}_{M1}$, $\mathcal{N}_{R1}$  & 95.4\% & 98.4\% & 98.2\% \\ 
    $\mathcal{N}_{M2}$, $\mathcal{N}_{R2}$  & 88.8\% & 94.8\% & 96.8\% \\ 
    $\mathcal{N}_{M3}$, $\mathcal{N}_{R3}$ & 83.4\% &  90.2\% & 92.1\% \\
  \end{tabular}
  \end{ruledtabular}
  %}

\end{table}

After setting the anomaly detection threshold, we compared the {\modelname}'s response to test EOSs with phase transitions. The detection efficiency results are shown in Tab.~\ref{tab:detection_efficiency}. Regardless of measurement uncertainties or the number of observations, only one $M(R)$ sequence based on $EOS_6$ (see right plot in Fig.~\ref{fig:eos} for details), was confidently detected with almost $100\%$ efficiency in almost all studied cases. We interpret this result as an obvious consequence of the $EOS_6$ $M(R)$ relation being the most different from the training dataset, e.g. the quark-matter branch extending towards large masses and radii. As for the other EOSs we observed the following interesting patterns. Models based on $EOS_{2-4}$, i.e. with an increasing average gravitational mass at which the phase transition occurs had gradually larger detection efficiencies at FPR=1\%. The pattern was observed in both cases of increased number of observations as well as presence of measurement uncertainties. However, the {\modelname} struggled to detect anomalies in $EOS_1$ samples even with no measurement uncertainties. $EOS_1$ represents a case for a low-mass phase transition $M(R)$ curve, therefore to a large extent, it resembles the $M(R)$ relations for training EOSs without phase transition. Additionally, the realistic number of observations assumed in the study is insufficient to properly probe the low-mass region of the $M(R)$ with the assumed NS mass function (Sect.~\ref{sect:meth_data}) in the case of this EOS. The same problem is even more evident for $EOS_2$ for the case of the largest uncertainties, and for some extent for the $EOS_5$ case (high-mass ``twin stars'' phase transition), for which the second $M(R)$ branch corresponds to a narrow range of masses. These results, in our opinion, reflect the importance of the NS mass function in statistical inference studies of this type.  

\begin{table*}
\caption{Summary of detection efficiency at FPR=$1\%$ for all studied EOSs with phase transitions in terms of adopted measurement uncertainties, as well as the number of observations $N$. For the discussion of results, see the text.}
\label{tab:detection_efficiency}
  %\centering
  %\resizebox{\textwidth}{!}{
  \begin{ruledtabular}
  \begin{tabular}{ccccccc} 
    $M(R)$ input data & $EOS_{1}$ & $EOS_{2}$ & $EOS_{3}$ & $EOS_{4}$ & $EOS_{5}$ & $EOS_{6}$ 
    \\  \hline  
    No error, $N = 10$ & 30.3\% & 76.6\% & 92.1\% & 99.9\% & 92.8\% & 100.0\% \\ 
    $\mathcal{N}_{M1}$, $\mathcal{N}_{R1}$, $N= 10$ & 11.8\% & 67.4\% & 86.1\% & 99.7\% & 56.1\% & 99.9\% \\ 
    $\mathcal{N}_{M2}$, $\mathcal{N}_{R2}$, $N= 10$ & 6.9\% & 10.8\% & 34.9\% & 62.6\% & 37.2\% & 100.0\% \\ 
    $\mathcal{N}_{M3}$, $\mathcal{N}_{R3}$, $N= 10$ & 8.7\% & 3.5\% & 18.3\% & 23.4\% & 21.9\% & 99.1\% \\ \hline
    
    No error, $N = 30$ & 28.7\% & 89.3\% & 99.7\% & 100.0\% & 87.6\% & 100.0\% \\ 
    $\mathcal{N}_{M1}$, $\mathcal{N}_{R1}$, $N = 30$ & 15.2\% & 86.8\% & 98.5\% & 100.0\% & 92.5\% & 100.0\% \\ 
    $\mathcal{N}_{M2}$, $\mathcal{N}_{R2}$, $N = 30$ & 13.8\% & 44.0\% & 80.8\% & 97.5\% & 46.9\% & 100.0\% \\ 
    $\mathcal{N}_{M3}$, $\mathcal{N}_{R3}$, $N = 30$ & 19.0\% & 11.1\% & 41.8\% & 49.8\% & 24.8\% & 100.0\% \\ \hline 
    
    No error, $N = 50$ & 38.8\% & 97.0\% & 99.9\% & 100.0\% & 97.3\% & 100\% \\ 
    $\mathcal{N}_{M1}$, $\mathcal{N}_{R1}$, $N = 50$ & 9.2\% & 90.5\% & 99.5\% & 100.0\% & 82.5\% & 100.0\% \\ 
    $\mathcal{N}_{M2}$, $\mathcal{N}_{R2}$, $N = 50$ & 16.5\% & 65.5\% & 92.7\% & 99.7\% & 52.6\% & 100.0\% \\ 
    $\mathcal{N}_{M3}$, $\mathcal{N}_{R3}$, $N = 50$ & 22.3\% & 17.6\% & 55.8\% & 68.7\% & 29.2\% & 100.0\% \\ 
  \end{tabular}
  \end{ruledtabular}
  %}
\end{table*}

%________________________________________________________________
\section{Conclusions} 
\label{sect:conclusions} 
%________________________________________________________________

We presented here a novel approach based on the AD technique enhanced by NF ({\modelname}), to the dense-matter EOS phase transition detection, which consists of analyzing the functionals of the EOS (NS astrophysical parameters: masses and radii) in order to search for characteristic signatures of the dense-matter phase-transitions. Specifically, we study how EOSs featuring various examples of phase transitions reflect on the $M(R)$ distributions and affect the latent representation of the {\modelname} model. In order to quantify our findings, we introduce the Euclidean distance, Eq.~\ref{eq:eucl_distance}, between the ``standard'' latent space distribution (featuring no phase transition) and the distribution under test, as a metric for setting the anomaly detection threshold. The threshold was associated with a specific value of FPR, namely $1\%$, and was calculated using ROC curves.

In Table~\ref{tab:detection_efficiency} we detailed how the number of observations and size of measurement uncertainties affected the {\modelname} performance. The latter has a significantly stronger effect on the final results in terms of AUC or detection efficiency for FPR=$1\%$. The results could be improved by increasing the number of observations, but the effect improves the results only partially.

Because of its outstanding character in comparison to the training set, $EOS_6$ \cite{2014PhRvD..89d3014D} was confidently detected as an outlier in all cases. However, the {\modelname} under-performed in the case of $EOS_1$ and $EOS_2$, which feature a low mass at which the phase transition occurs, as well as, to some extend, in the case of $EOS_5$ (high mass phase transition), for which a range of masses of the high-density $M(R)$ branch is very narrow. These results reflect the variable amount of information supplied to the {\modelname} through the selection of $M(R)$ observables by means of the NS mass function, i.e. for these EOSs the phase-transition mass range is not sampled sufficiently well to provide enough information to the {\modelname} model, with given number of observations and assumed NS mass function.  However, for the $\mathcal{N}_{M2}$, $\mathcal{N}_{R2}$ error probability distributions with $\sigma_{M_2}=0.05\,M_\odot$ and $\sigma_{R_2}=0.5\,M_\odot$, and a moderate size data sample (N=30), we demonstrate that the signature of detached $M(R)$ branches is robustly detected if the phase transition point is sampled sufficiently well by the observations, as in the cases of $EOS_3$ and $EOS_4$.

Inconclusive results obtained from the real $M(R)$ measurements in Appendix~\ref{sect:real_data}, reflect current state-of-the-art of the quality of observations: to be able to make conclusive statements on the existence (or lack) of strong dense-matter phase transitions from the $M(R)$ observations, one needs data with significantly smaller $M$ and $R$ uncertainties, as well as a larger observational sample. 

A study pertaining to other observables, e.g. the values obtained in the GW inspiral events \cite{2021arXiv211103606T}, such as the chirp mass and tidal deformability, is a viable possibility for a follow-up study of the {\modelname} capabilities, given the planned increase of sensitivity of GW detectors and expected number of high signal-to-noise events, which will at some moment exceed the number of EM measurements \cite{2018LRR....21....3A}.

\begin{acknowledgements}
This work was partially supported by the Polish National Science Centre grants 2016/22/E/ST9/00037, 2017/26/M/ST9/00978, 2020/37/N/ST9/02151 and 2021/43/B/ST9/01714, Polish National Agency for Academic Exchange grant PPN/IWA/2019/1/00157 as well as the European Cooperation in Science and Technology COST Action G2net (no. CA17137). The Quadro P6000 GPU used in this research was donated by the NVIDIA Corporation.
The authors would like to express their gratitude to Profs. Ik Siong Heng and Chris Messenger of the University of Glasgow for their inspiring ideas about using normalizing flows as a potentially powerful method for detecting anomalies in astrophysical data, and the anonymous referee for their constructive comments. 
\end{acknowledgements}

\appendix
\section{Evaluation on real astrophysical measurements} 
\label{sect:real_data} 

\begin{figure}[ht]
    \vspace{10pt} 
    \includegraphics[width=\columnwidth]{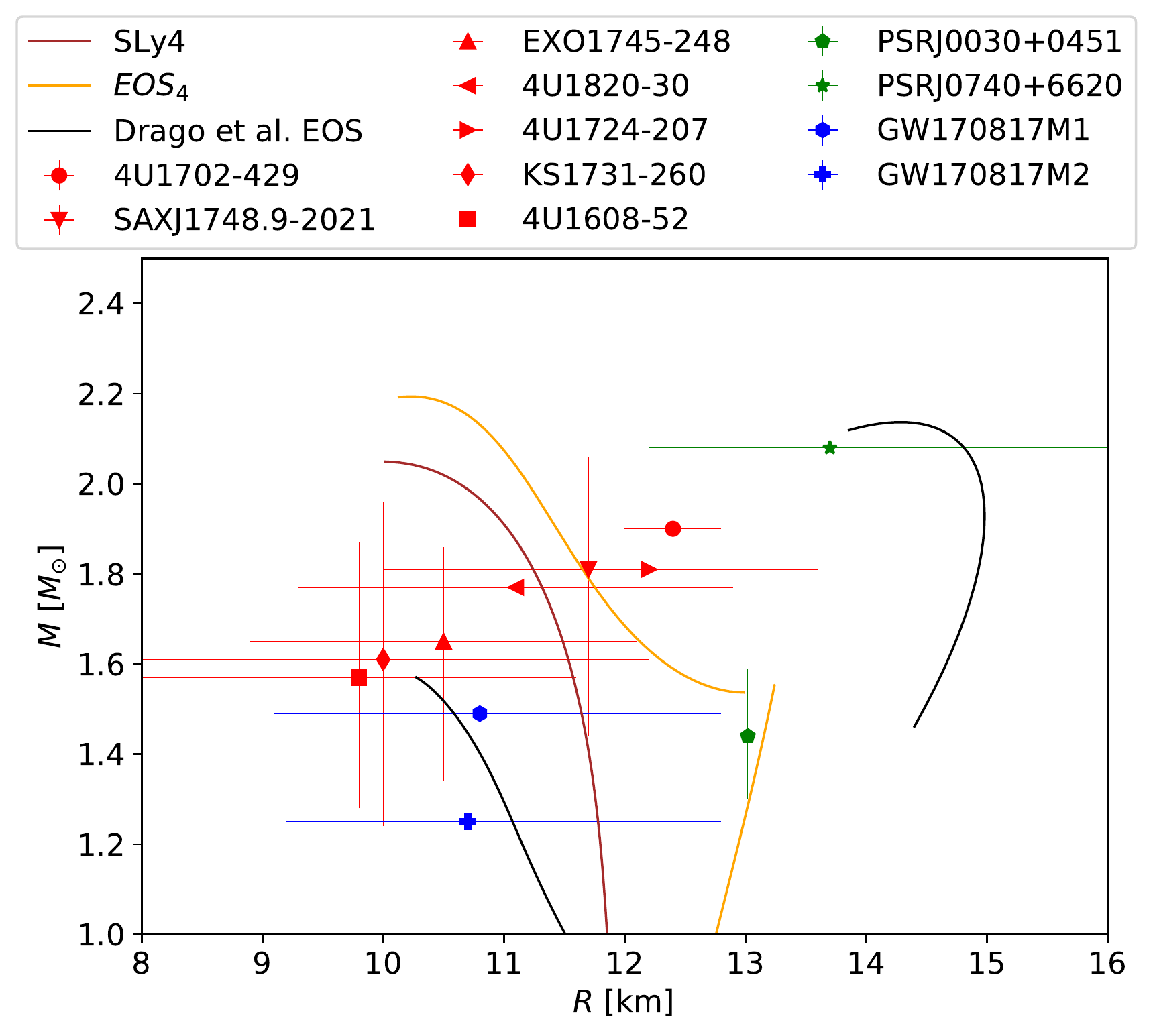}
    \caption{A set of real $M(R)$ measurements corresponding to observed NSs. Red symbols denote X-ray bursters/Low Mass X-ray Binaries \cite{Ozel_2016,2016ARA&A..54..401O,2017A&A...608A..31N}, green symbols denote NICER sources \cite{2019ApJ...887L..24M,2021ApJ...918L..28M}, blue symbols denote GW170817 $M(R)$ estimates \cite{2017PhRvL.119p1101A,2018PhRvL.121p1101A,2019PhRvX...9a1001A}. For comparison, the SLy4 EOS $M(R)$ (brown curve), $EOS_4$ $M(R)$ (orange curves) and $EOS_6$ $M(R)$ (black curves) are plotted. All error bars are 68\% uncertainty, except the case of GW170817 (see text for details). The mass and radius ranges are as in Fig.~\ref{fig:eos}.}
    \label{fig:mr-realdata}
\end{figure}

Once trained and tested against simulated data, the model was exposed to real data available at the time. We have compiled a list of simultaneous measurements of NS mass $M$ and radius $R$, associated with $11$ different NSs. To our best knowledge, this list represent the state of art of the field. The objects and their $M(R)$ measurements shown in Fig.~\ref{fig:mr-realdata} are as follows (errors denoting 68\% uncertainty, unless stated otherwise): thermonuclear X-ray burster 4U1702-429 with $M = 1.9{\pm}0.3\,M_\odot$, $R = 12.4{\pm}0.4$ km \cite{2017A&A...608A..31N}, Low Mass X-ray Binaries \cite{Ozel_2016,2016ARA&A..54..401O}: SAXJ1748.9-2021 ($M=1.81^{+0.25}_{-0.37}\,M_\odot$, $R=11.7{\pm}1.7$ km), EXO1745-248 ($M=1.65^{+0.21}_{-0.31}\,M_\odot$, $R=10.5{\pm}1.6$ km), 4U1820-30 ($M=1.77^{+0.25}_{-0.28}\,M_\odot$, $R=11.1{\pm}1.8$ km), 4U1724-207 ($M=1.81^{+0.25}_{-0.37}\,M_\odot$, $R=12.2{\pm}1.4$ km), KS1731-260 ($M=1.61^{+0.35}_{-0.37}\,M_\odot$, $R=10.0{\pm}2.2$ km), 4U1608-52 ($M=1.57^{+0.30}_{-0.29}\,M_\odot$, $R=9.8{\pm}1.8$ km), NICER sources PSRJ0030+0451 ($M=1.44^{+0.15}_{-0.14}\,M_\odot$, $R=13.02^{+1.24}_{-1.06}$ km, \cite{2019ApJ...887L..24M}) and PSRJ0740+6620 ($M=2.08^{+0.07}_{-0.07}\,M_\odot$, $R=13.7^{+2.6}_{-1.5}$ km, \cite{2021ApJ...918L..28M}), as well as inferred radius measurement of the GW170817 event components ($M_1=1.48^{+0.12}_{-0.12}\,M_\odot$, $R_1=10.8^{+2.0}_{-1.7}$ km, $M_2=1.26^{+0.1}_{-0.1}\,M_\odot$, $R_2=10.7^{+2.1}_{-1.5}$ km, where mass values are arithmetic averages from ranges given in \cite{2017PhRvL.119p1101A,2019PhRvX...9a1001A} for a low-spin prior case, and radius values are 90\% credible level EOS-insensitive estimates based on GW detectors' data alone, \cite{2018PhRvL.121p1101A}).   

The {\modelname} did not reliably indicate presence of dense-matter phase-transition anomalies in this dataset, as the Euclidean distance metric computed was below the assumed detection threshold.  For example, when tested on the {\modelname} trained on dataset with $\mathcal{N}_{R3}$, $\mathcal{N}_{M3}$ uncertainties and $N=10$ observations, the corresponding Euclidean distance $D$ for the randomly taken $10$ real observations (the {\modelname} required input of the fixed size) was $D=5.2$, a value below the anomaly detection threshold $D=5.6$ for these uncertainties and $N$. In our tests, the values of resulting $D$ value changed depending on how the model was trained (on more or less ``noisy'' data), but the conclusion remained the same: the Euclidean distance $D$ corresponding to the real data was below the anomaly detection threshold of the {\modelname}. This negative result is understandable by  comparing the magnitude of real data errors in $M$ and $R$ with the assumed ``worst quality'' simulated error probability distributions $\mathcal{N}_{M3}$, $\mathcal{N}_{R3}$  ($\sigma_{M_3}=0.1\,M_\odot$, $\sigma_{R_3}=1\,M_\odot$), and the results of {\modelname} on a similar size sample (N=10), gathered in Table~\ref{tab:detection_efficiency}.

We cannot claim however, that this result constitutes a significant statement of the {\em lack of existence} of dense-matter phase transitions in the data either. Given the heterogeneous nature of the data (potentially existing unknown and/or complicated systematics), and the fact that the measurement uncertainties in the real dataset are large (much larger than those used for training and testing, e.g. assumed in $\mathcal{N}_{M3}$ and $\mathcal{N}_{R3}$ distributions), we interpret this result as inconclusive: the level of measurement accuracy as well as the number of $M(R)$ measurements does not allow, at the present moment, to produce a meaningful statement of the existence (or not) of the dense-matter phase transition signature in the astrophysical data.

\bibliography{bibfile} % Produces the bibliography via BibTeX.

\end{document}